\documentclass{aa}  

\usepackage{graphicx}
\usepackage{hyperref}
\usepackage[hyphenbreaks]{breakurl}

\usepackage{txfonts}
\begin{document}

   \title{The XMM-Newton serendipitous survey\thanks{Based on observations obtained with XMM-Newton, an ESA science mission with instruments and contributions directly funded by ESA Member States and NASA.}}

   \subtitle{IX. The fourth XMM-Newton serendipitous source catalogue}

   \author{N. A. Webb
          \inst{1}
          \and
          M. Coriat\inst{1}
          \and
          I. Traulsen\inst{2} 
          \and
          J. Ballet\inst{3}
          \and
          C. Motch\inst{4}
          \and
          F. J. Carrera\inst{5}
          \and
          F. Koliopanos\inst{1}
          \and
          J. Authier\inst{1}
          \and
          I. de la Calle\inst{6}
          \and
         M. T. Ceballos\inst{5}
          \and
         E. Colomo\inst{6}
         \and 
          D. Chuard\inst{7,3}
          \and
          M. Freyberg\inst{8}
          \and
          T. Garcia\inst{1}
          \and
          M. Kolehmainen\inst{4}
          \and
          G. Lamer\inst{2}
           \and
          D. Lin\inst{9}
          \and
          P. Maggi\inst{4}
          \and
          L. Michel\inst{4}
           \and
          C. G. Page\inst{10}
           \and
           M. J. Page\inst{11}
          \and
          J. V. Perea-Calderon\inst{12}
          \and
          F.-X. Pineau\inst{4}
          \and
         P. Rodriguez\inst{6}
          \and
          S.R. Rosen\inst{6}
         \and
          M. Santos Lleo\inst{6}
          \and
          R. D. Saxton\inst{6}
          \and
          A. Schwope\inst{2}
          \and
          L. Tom\'as\inst{6}
           \and
          M. G. Watson\inst{10}
          \and
          A. Zakardjian\inst{1}
                   }

   \institute{IRAP, Universit\'e de Toulouse, CNRS, CNES, Toulouse, France\\
              \email{Natalie.Webb@irap.omp.eu}
          \and
Leibniz-Institut für Astrophysik Potsdam (AIP), An der Sternwarte 16, 14482 Potsdam, Germany
          \and
IRFU, CEA, Universit\'e Paris-Saclay, F-91191 Gif-sur-Yvette, France 
          \and
Universit\'e de Strasbourg, CNRS, Observatoire astronomique de Strasbourg, UMR 7550, 67000 Strasbourg, France
          \and
Instituto de Física de Cantabria (CSIC-UC), Avenida de los Castros, 39005 Santander, Spain
         \and
ESAC, European Space Astronomy Center (ESAC-ESA), Madrid 28691,  Spain 
\and
Universit\'e de Paris, CNRS, Astroparticule et Cosmologie, F-75013 Paris, France
         \and       
Max-Planck-Institut f\"ur extraterrestrische Physik, Giessenbachstra{\ss}e 1, 85748 Garching, Germany
          \and
Space Science Center, University of New Hampshire, Durham, NH, USA
          \and
Department of Physics \& Astronomy, University of Leicester, Leicester, LE1 7RH, UK
          \and
Mullard Space Science Laboratory, University College London, Holbury St Mary, Dorking, Surrey RH5 6NT, UK
          \and
RHEA for ESA/ESAC. European Space Astronomy Center (ESAC-ESA). Madrid. 28691. Spain
}
 
   \date{Received , accepted }

% \abstract{}{}{}{}{} 
% 5 {} token are mandatory
 
  \abstract
   {Sky surveys produce enormous quantities of data on extensive regions of the sky. The easiest way to access this information is through catalogues of standardised data products. {\em XMM-Newton} has been surveying the sky in the X-ray, ultra-violet, and optical bands for 20 years. }
   {The {\em XMM-Newton} Survey Science Centre has been producing standardised data products and catalogues to facilitate access to the serendipitous X-ray sky.}
   {Using improved calibration and enhanced software, we re-reduced all of the 14041 {\em XMM-Newton} X-ray observations, of which 11204 observations contained data with at least one detection and with these we created a new, high quality version of the {\em XMM-Newton} serendipitous source catalogue, 4XMM-DR9. }
   {4XMM-DR9 contains 810795 detections  down to a detection significance of 3 $\sigma$, of which 550124 are unique sources, which cover 1152 degrees$^{2}$ (2.85\%) of the sky.  Filtering 4XMM-DR9 to retain only the cleanest sources with at least a 5 $\sigma$ detection significance leaves 433612 detections. Of these detections, 99.6\% have no pileup. Furthermore, 336 columns of information on each detection are provided, along with images. The quality of the source detection is shown to have improved significantly with respect to previous versions of the catalogues. Spectra and lightcurves are also made available for more than 288000 of the brightest sources (36\% of all detections). }
   {}


   \maketitle
%
%-------------------------------------------------------------------

\section{Introduction}
%- NW

The sky is constantly being surveyed by many different telescopes exploiting the full range of the electromagnetic spectrum, in addition to gravitational wave, neutrino, and cosmic ray observatories. Each observation can provide a clue as to the nature of the source and the physical processes underway. In addition, many objects are known to be highly variable in time, requiring many observations to fully understand the nature of the variability. Whilst dedicated observations can be necessary to answer some science questions, frequently, catalogues can provide the required information. Catalogues can also provide homogeneous datasets for classes of objects as well as reveal previously unknown objects. 

Catalogues have been produced for the majority of the X-ray missions that have flown. Early X-ray missions detected very few objects. The fourth version of the UHURU catalogue \citep[1970-1973,][]{form78} indicates that just 339 X-ray sources were discovered by the satellite. The HEAO 1 catalogue \citep[1977-1978,][]{wood84} provides 842 X-ray sources. The ROSAT catalogue, 2RXS \citep[1990-1991,][]{boll16} gives 135000 X-ray detections or 129192 sources. However, more recent X-ray observatories have several advantages over the earlier missions. Firstly, they have a larger collecting area and are therefore more sensitive. Secondly, they have also surveyed the sky for a much longer period and thus they detect many more sources. Chandra, which was launched in July 1999, boasts a very extensive catalogue, the Chandra Source Catalog Release 2.0 (CSC 2.0) \citep{evans2014,chen19} with 928280 X-ray detections, which are from 317167 individual X-ray sources. The Neil Gehrels Swift Observatory was launched in November 2004 and the 2SXPS catalogue \citep{evan20} lists 1.1 million detections, which are of 206335 individual X-ray sources. The major advantage of this catalogue is that it covers a large field of view,  that is 3790 deg$^2$ of sky, and sources have been pointed many times over the last 16 years. The first catalogue of sources detected with the hard X-ray observatory, NuSTAR, lists 497 sources after 40 months of observations \citep{lans17}.

This paper focuses on the catalogue of detections from the European Space Agency's second cornerstone mission from the Horizon 2000 programme, {\em XMM-Newton} \citep{jans01}, which was launched twenty years ago on 10 December 1999. It has the largest effective area of any
X-ray satellite \citep{ebre19} thanks to the three X-ray telescopes aboard, each with $\sim$1500 cm$^2$ of geometric effective area. This fact, coupled with the large field of view (FOV) of 30\arcmin\ diameter, means that a single pointing with the mean duration in the catalogue of 37 ks detects 70-75 serendipitous X-ray sources. The catalogue of serendipitous sources from overlapping XMM-Newton observations 4XMM-DR9s is described in paper X of this series (Traulsen et al., accepted).

The XMM-Newton Survey Science Centre\footnote{\url{http://xmmssc.irap.omp.eu/}} (SSC), a consortium of ten European Institutes \citep{wats01}, has developed much of the XMM-Newton Science Analysis System (SAS) \citep{sas04} which reduces and analyses {\em  XMM-Newton} data and created pipelines to perform standardised routine  processing of the {\em XMM-Newton} science data. The XMM-SSC also produces catalogues of all of the detections made with {\em XMM-Newton}. The catalogues of X-ray detections made with the three EPIC \citep{stru01,turn01} cameras that are placed at the focal point of the three X-ray telescopes have been designated 1XMM, 2XMM, and 3XMM \citep{wats09}, with incremental versions of these catalogues indicated by successive data releases, denoted -DR in association with the catalogue number. This paper presents the latest version of the XMM catalogue, 4XMM, which spans 19 years of observations made with {\em XMM-Newton} and includes many improvements with respect to previous {\em XMM-Newton} catalogues. The most notable change between 3XMM and 4XMM is the methodology used for background modelling (see Sec.~\ref{sec:bkg}).

%--------------------------------------------------------------------
\section{Catalogue observations}
% - NW
\label{sec:catobs}

\begin{figure}
   \centering
   \includegraphics[width=9cm]{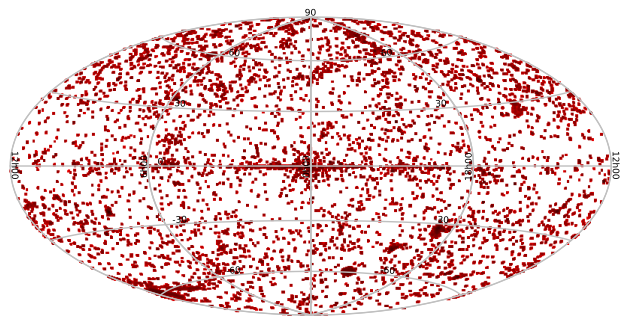}
      \caption{Hammer-Aitoff equal area projection in Galactic coordinates of
the 11204 4XMM-DR9 fields.
              }
         \label{fig:hammer_aitoff}
   \end{figure}

\begin{figure}
   \centering
   \includegraphics[width=9.5cm]{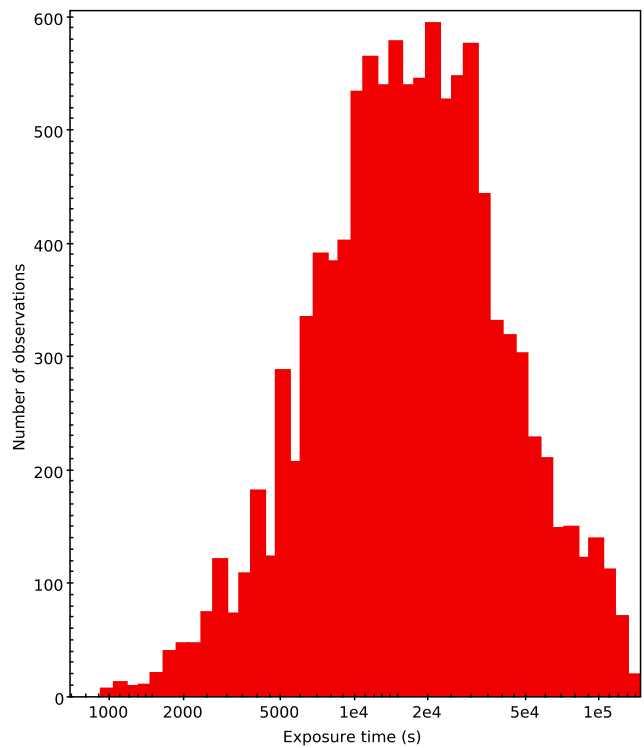}
      \caption{Distribution of MOS 1 good exposure time (after event filtering) for the observations included in the 4XMM-DR9 catalogue.
              }
         \label{fig:M1ExpTime}
   \end{figure}

A total of 14041 XMM-Newton EPIC observations were publicly available as of 1 March 2019, but only 11204 of these observations had at least one detection.  4XMM-DR9 is made from the detections drawn from the 11204 {\em XMM-Newton} EPIC observations. The repartition of data modes for each camera and observation can be found in Table~\ref{tab:11204obs}. The Hammer-Aitoff equal area projection in Galactic coordinates of
the 4XMM-DR9 fields can be seen in Fig.~\ref{fig:hammer_aitoff}.  All of those observations containing $>$~1 ks clean data ($>$1 ks
of good time interval for the combined EPIC exposure) were retained for the
catalogue. Fig.~\ref{fig:M1ExpTime} shows the distribution of
total good exposure time (after event filtering) for the observations included
in  the 4XMM-DR9 catalogue and using any of the thick, medium or thin filters, but not the open filter. Open filter data were processed but not used in the
source detection stage of pipeline processing. The same {\em XMM-Newton} data 
modes were used as in 2XMM \citep{wats09} and are included in Table~\ref{tab:datamodes} of this paper, for convenience. The data in 4XMM-DR9
include 322 observations that were publicly available at the time of creating
3XMM-DR8, but were not included in that version due to high background or processing problems. Due to changes in the pipeline and in the background modelling, these problems have been overcome and thus the data could be included in 4XMM-DR9.

\begin{table*}
      \caption[]{Characteristics of the 11204 {\em XMM-Newton} observations included
in the 4XMM-DR9 catalogue.}
         \label{tab:11204obs}
   \begin{tabular}{lccccccc}     % 8 columns 
\hline\hline       
                      % To combine 4 columns into a single one 
Camera & \multicolumn{3}{c}{Modes} & \multicolumn{3}{c}{Filters} & Total\\ 
 & Full$^a$ & Window$^b$ & Other$^c$ & Thin & Medium & Thick & \\
\hline                    
pn & 8462 & 683 & 1344 & 5640 & 4011 & 838 & 10489 \\
MOS1 & 8681 & 1950 & 373 & 5080 & 4943 & 981 & 11004 \\
MOS2 & 8728 & 1981 & 348 & 5120 & 4971 & 966 & 11057 \\
\hline                  
\end{tabular}
\\ $^a$ Prime Full Window Extended (PFWE) and Prime Full Window (PFW) modes;
$^b$ pn Prime Large Window (PLW) mode and any of the various MOS
Prime Partial Window (PPW) modes; $^c$  other pn modes such as the Small Window, timing or burst modes, MOS modes (Fast Uncompressed (FU), Refresh Frame Store (RFS)).
   \end{table*}

\begin{table}[t]
\normalsize
\caption{Data modes of XMM-Newton exposures included in the 4XMM catalogue.} 
\label{tab:datamodes}
\small
\centering
\begin{tabular}{lll}
\hline \hline
Abbr. & Designation & Description \\
\hline
\multicolumn{3}{l}{\it \ \ MOS cameras:} \\
PFW  &	Prime Full Window & covering full FOV \\
PPW2 &	Prime Partial W2 & small central window \\
PPW3 &	Prime Partial W3 & large central window \\
PPW4 &	Prime Partial W4 & small central window \\
PPW5 &	Prime Partial W5 & large central window \\
FU &	Fast Uncompressed & central CCD in timing mode \\
RFS &   Prime Partial RFS  & central CCD with different frame \\
    &                      & \ \ time (`Refreshed Frame Store') \\
\multicolumn{3}{l}{\it \ \ pn camera:} \\
PFWE  &	Prime Full Window & covering full FOV \\
      &	 \ \ \ Extended & \\
PFW &	Prime Full Window & covering full FOV  \\
PLW &	Prime Large Window & half the height of PFW/PFWE \\[0.2cm]
\hline 
\end{tabular}
\normalsize		
\end{table}

\section{Data processing}
% - NW
\label{sec:processing}

Data processing for the 4XMM-DR9 catalogue was based on the SAS version 18 and carried out with the pipeline version 18\footnote{\url{https://www.cosmos.esa.int/web/xmm-newton/pipeline-configurations}} and the latest set of current calibration files at the time of processing (February and March 2019). 

The main data processing steps used to produce the 4XMM data products were
similar to those outlined in \cite{rose16,wats09} and described on the SOC
 webpages\footnote{\url{https://xmm-tools.cosmos.esa.int/external/xmm_products/pipeline/doc/17.40_20181123_1545/modules/index.html}}.  For all the 4XMM data, the observation data files were processed to produce calibrated event lists. The optimised background time intervals were identified and using them, the filtered exposures (taking into account exposure time, instrument mode, etc.), multi-energy-band X-ray images, and
exposure maps were generated. The source detection was done simultaneously on all images and bands, 1$-$5, from the three cameras as in \cite{wats09,rose16}.  The probability, and corresponding likelihood, were computed from the null hypothesis that the measured counts in the search box result from a Poissonian fluctuation in the estimated background level.  A detection mask was made for each camera that defines the area of the detector which is suitable for source detection. An initial source list was made using a ‘box  detection’ algorithm. This slides a search box  (20\arcsec\ $\times$ 20\arcsec) across the image defined by the detection mask. Sources were cut-out using a radius that was dependent on source brightness in each band, and these areas of the image where sources had been detected were blanked out. The source-excised images, normalised by the exposure maps, and the corresponding masks are convolved with a Gaussian kernel to create the background map \citep[see][where this smoothing method is new for the detection catalogue]{trau19}.  A second box-source-detection pass was then carried out, creating a new source list, this time using the background maps (‘map mode’) which increased the source detection sensitivity compared to the first pass.  The box size was again set to 20\arcsec\ $\times$ 20\arcsec. A maximum likelihood fitting procedure was then applied to the sources to calculate source parameters in each input image, by fitting a model to the distribution of counts over a circular area of radius 60\arcsec, see \cite{wats09}.  For the catalogue of detections (4XMM-DR9), source  parameterisation was done before  cross-correlation of the source list with a variety of
archival catalogues, image databases, and other archival resources. The creation of spectra and light curves for the brightest sources was then carried out.  Automatic and visual screening procedures were carried out to check for any problems in the data products.  The data from this processing have been made available through the {\em XMM-Newton} Science Archive\footnote{\url{https://nxsa.esac.esa.int/nxsa-web}} (XSA), but see also Sec.~\ref{sec:access}.

%For the stacked catalogue (4XMM-DR9s), the ensuing procedure is described in Traulsen et al. (submitted). It is very similar to the procedure used to produce the catalogue of detections, except that the source detection is carried out on all overlapping observations simultaneously and the results per input image are stored in an intermediate source list, before determining the final source list and parameterisation, see also Sec.~\ref{sec:stackedcat}. The catalogue is cross-matched with 4XMM-DR9, and long-term lightcurves and auxiliary images of the sources are produced.

\subsection{Exposure selection}
% - NW
\label{sec:expsel}

The same criteria used for selecting exposures for 3XMM were retained for 4XMM. 
A total exposure time of 410 Ms was available for 4XMM-DR9, with an increase of 57\% compared to 3XMM-DR5.
%Almost 50\% more exposures (or 10.7 Gs) were available for 4XMM-DR9 compared to 3XMM-DR5. 

\subsection{Event list processing}
% - NW
\label{sec:evliproc}

Much of the pipeline processing that converts raw ODF event file data from the
EPIC instruments into cleaned event lists has remained unchanged from the
pre-cat9.0 pipeline and is described in section 4.2 of
\cite{wats09}. A number of improvements have been made since the 2XMM \citep{wats09} and 3XMM \citep{rose16} catalogues, which can be found in the SAS release notes\footnote{\url{https://www.cosmos.esa.int/web/xmm-newton/sas-release-notes/}}. These include source spectra and light curves created for pn Timing mode and small window data, source detection on pn small window data, energy dependent Charge Transfer Inefficiencies (CTI) and double event energy corrections, time and pattern dependent corrections of the spectral energy resolution of pn data, X-ray loading and rate dependent energy (PHA) and CTI corrections for EPIC pn Timing and Burst modes, binning of MOS spectra changed from 15 eV to 5 eV. Filtering was carried out with XMMEA\_EM, which is a bit-wise selection expression, automatically removing bad events such as bad rows, edge effects, spoiled frames, cosmic ray events (MIPs), diagonal events, event beyond threshold, etc, instead of XMMEA\_SM (which removed all flagged events except those flagged only as CLOSE\_TO\_DEADPIX). Other modifications include the generation of background regions for EPIC spectra and light curves selected from the same EPIC chip where the source is found, observations of solar system objects processed such that X-ray images and spectra correctly refer to the moving target, the inclusion of pileup diagnostic numbers for EPIC sources (see also Sec.~\ref{sec:pileup}), and footprints for EPIC observations based on combined EPIC exposure maps provided as ds9 region files. Other changes carried out specifically for the production of 4XMM include a revised systematic position error (see Sec.~\ref{sec:poserr}), the modelling of the EPIC background (see Sec.~\ref{sec:bkg}) and finer binning of EPIC lightcurves (see Sec.~\ref{sec:lcs}). A small rotation of $\sim$0.4$^\circ$ was noted in 3XMM fields, but analysis of 4XMM data shows that the recent improvements to calibration have resolved this issue. Below we describe some of the more recent developments specifically implemented for 4XMM.

\subsection{Systematic position error}
% - FC
\label{sec:poserr}

The astrometry of the X-ray detections is improved by using the catcorr task to cross-correlate the X-ray detections with the USNO B1.0, 2MASS or SDSS (DR8) optical or IR catalogues. Using pairs of X-ray and optical or infra-red detections that fall within 10\arcsec\ of each other, the astrometry for the field is corrected using a translational shift in the right ascension (RA) and declination (DEC) directions, together with the rotational error component. A systematic error on the position (SYSERRCC) is then calculated using the 1 $\sigma$ errors on the shifts in the RA ($\Delta\alpha_{error}$) and DEC ($\Delta\delta_{error}$) directions and the rotational error component in radians ($\Delta\theta_{error}$), derived from from the catalogue that yields the 'best' solution, using $SYSERRCC = \sqrt{(\Delta\alpha_{error}^2 + \Delta\delta_{error}^2 + (r * \Delta\theta_{error})^2)}$, where $r$ is the radial off-axis angle of the detection from the spacecraft boresight in arcsecs. However, where catcorr fails to obtain a statistically reliable result (poscorrok=false), a systematic error of 1.5\arcsec\ was used to create the 3XMM catalogue. 

In the framework of creating 4XMM, this systematic error was re-evaluated. In order to determine an improved systematic error, we identified fields in 3XMM-DR8 where  catcorr failed. We used sources from the Sloan Digital Sky Survey Data Release 12 quasar (SDSS~DR12~QSO) catalogue \citep{Paris17} with good quality spectra (\texttt{ZWARNING=0}) and point-like morphology (\texttt{SDSS\_MORPHO=0}). To avoid mismatches between targets and matched photometry\footnote{see \url{https://www.sdss.org/dr12/algorithms/match/}}  we chose non-empty \texttt{OBJ\_ID} values. We then cross-matched with the SDSS~DR9 photometry catalogue \citep{SDSSDR9}  in Vizier\footnote{\url{http://cdsarc.u-strasbg.fr/viz-bin/cat/V/139}} with a maximum distance of 5\arcsec. This step provided the uncertainty in the astrometric position of SDSS. We adopted the radially-averaged uncertainty in the SDSS positions to which we had already added a systematic 0.1\arcsec\ in quadrature, $\Delta S=\sqrt{(\Delta\alpha^2+\Delta\delta^2)/2+0.1^2}$. We then discarded all quasars with more than one SDSS DR9 counterpart within 5~arcsec. Out of the 256107 ``clean'' quasars, we selected the potential counterparts to the 3XMM~DR8 sources, but also discarded those which could be counterparts of more than one 3XMM~DR8 source. We used the ``slim'' catalogue for this purpose, since multiple detections of the same physical source appear only once.
%%% fjc : these definitions would perhaps be better placed somewhere else ?
The total positional error on each source in the slim catalogue is \texttt{SC\_POSERR}, calculated as the weighted average of the total positional errors \texttt{POSERR} of the individual detections. In turn, this is calculated as \texttt{POSERR}=$\sqrt{RADEC\_ERR^2 + SYSERRCC^2}$, where \texttt{RADEC\_ERR}$\equiv \sqrt{\Delta\alpha_X^2+\Delta\delta_X^2}$ ($\Delta\alpha_X$ and $\Delta\delta_X$ are the 1$\sigma$ uncertainties in the RA and Dec coordinates, respectively).
%%%fjc
We cross-matched the SDSS~DR9 positions of  ``clean'' QSOs with the positions of the sources in the slim catalogue out to a distance of $r=30$\arcsec. For each of the resulting pairs we estimated the combined positional error as $\sigma=\sqrt{\Delta S^2+\Delta X^2/2}$, where $\Delta X\equiv$\texttt{SC\_POSERR}
and discarded all quasars that had more than one counterpart out to $r/\sigma=6$, leaving 7205 suitable QSO (there were 26 QSO with more than one counterpart out to that limit). There were no pairs of quasars that corresponded to the same X-ray source. 

Since each instance of an X-ray source in the 3XMM-DR8 detection catalogue is an independent measurement, we cross-matched the sample of suitable quasars with the detection catalogue where poscorrok=false, out to $r=30$\arcsec\ again, filtering the latter with \texttt{SUM\_FLAG=0} and \texttt{EP\_EXTENT=0}, to keep only the cleanest sample of secure point-like X-ray sources. At this point we have 178 quasar-X-ray detection pairs.  As for the slim catalogue, we define the combined positional error as $\sigma=\sqrt{\Delta S^2+\Delta X^2/2}$, where 
%%% fjc
$\Delta X$=\texttt{RADEC\_ERR}
%%% fjc
and $x=r/\sigma$. Our final filtering retained only the 157 QSO-X-ray pairs with $x<5$.

\begin{figure}
\centerline{\includegraphics[width=0.49\textwidth]{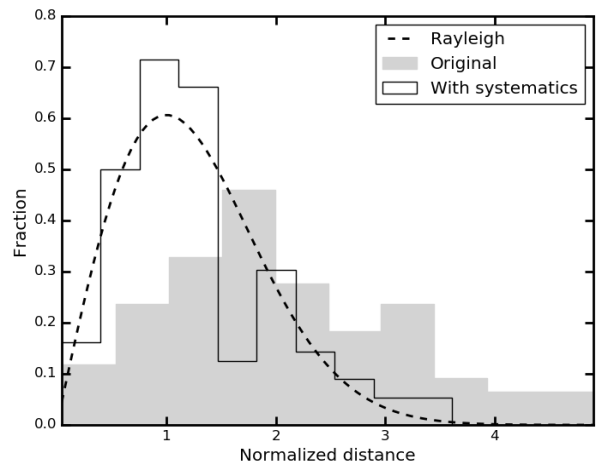}}
\caption{157 XMM-Newton-SDSS quasar pairs as a function of normalised distance $x$  before adding a systematic uncertainty (grey histogram) and after its addition (black solid line), along with the Rayleigh distribution (black dashed line).}
\label{fig:rayleigh157}
\end{figure}

The expected probability density distribution of $x$ should follow the Rayleigh distribution $P(x)=x e^{-x^2}$. Since this was not the case for the 157 pairs of sources found above, we added an additional positional uncertainty, $\Sigma$, in quadrature, so that the total positional uncertainty is now $\sigma'=\sqrt{\sigma^2+\Sigma^2}$, looking for the value of $\Sigma$ that minimises the difference between the distribution of the $x'\equiv r/\sigma'$ and the Rayleigh distribution
%%% fjc
using maximum likelihood.
%%%fjc
We found $\Sigma=1.29\pm 0.12$\arcsec, where the uncertainty (1$\sigma$) has been calculated by bootstrap with replacement. The improvement can be seen in Figure~\ref{fig:rayleigh157}. This value was then used to replace the 1.5\arcsec\ systematic error when  poscorrok=false. We note that a minor error was introduced into 4XMM-DR9, where the systematic error used in the case of poscorrok=false was a factor $\sqrt{2}$ too small. This is corrected in versions 4XMM-DR10 and higher.

\subsection{Modelling the EPIC background}
% - IT
\label{sec:bkg}

For each input image to the source detection, the background is modelled by an
  adaptive smoothing technique. The method was initially applied to the data in the 3XMM-DR7s catalogue which treats overlapping \textit{XMM-Newton} observations and is
  described by \cite{trau19}. Since 3XMM-DR7s was based on a
  selection of clean observations, the smoothing parameters were revised for
  the 4XMM catalogues, which cover observations of all qualities. The three
  parameters of the smoothing task are the cut-out radius to excise sources,
  the minimum kernel radius of the adaptive smoothing, and the requested
  signal-to-noise ratio in the map. Their best values were determined in a
  three-fold assessment which involved real observations, randomised images,
  and visual screening.

  656 observations were chosen which cover positions of cluster candidates
  identified by \citet{takey13} to involve a considerable number of extended
  and point-like sources. Their background was modelled using different
  combinations of the smoothing parameters, and source detection was
  performed. The number of detections and recovered clusters, and the
  source parameters of the clusters and point-like detections were
  compared, opting for a reasonable compromise between total number of
  detections and potentially spurious content and for reliable fluxes, and
  extent radius of the clusters. The source parameters of point-like
  detections were largely unaffected by the parameter choice in the tested
  parameter range.

  The optimisation was then re-run on ninety observations, in which the 
  background was replaced by a Poissonian randomisation. Finally, the two best
  combinations of smoothing parameters and the previously used spline fit were
  compared in a blind test. The detection images were inspected in randomised
  order, so the screeners could not know which source-detection results were
  based on which background model. The three parts of the assessment confirmed
  the preference for the adaptive smoothing approach over a spline fit and
  the estimation of the final parameters: a brightness threshold for the source cut-out radius of
  $2\times 10^{-4}\,\textrm{counts}\,{\rm arcsec}^{-2}$, a minimum smoothing
  radius of 10\,pixels (40\arcsec\ in default image binning), and a
  signal-to-noise ratio of 12.

\subsection{Updated flagging procedures}
\label{sec:flagging}

A single change to the flags provided for each detection has been introduced. Flag 12 now indicates if the detection falls on a region of the detector that can show hot pixels that can be misinterpreted as a source. Further information is provided in Section~\ref{sec:hotflag}.

\subsubsection{Hot areas in the detector plane}
% - JB
\label{sec:hotflag}

\begin{figure}[t]
\centering
\includegraphics[width=\linewidth]{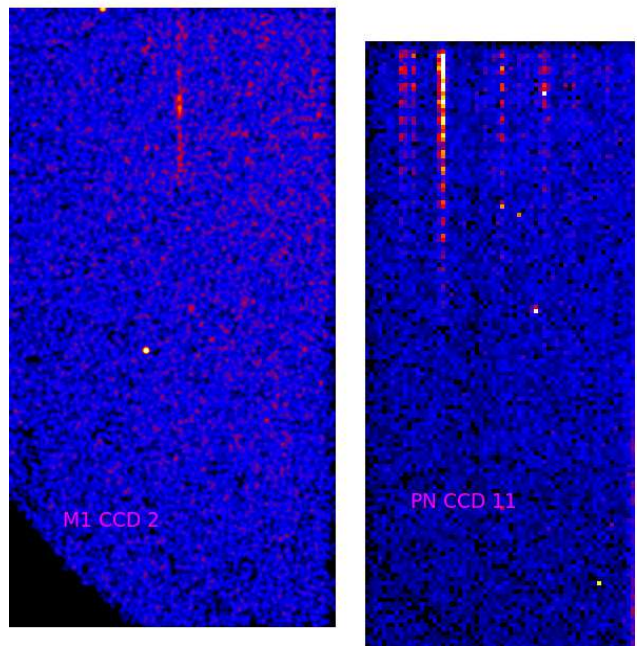}
\caption{Same part of the focal plane (lower left in pn detector coordinates) viewed by pn (CCD 11) and MOS1 (CCD 2). The maps are in CCD coordinates, but offset and zoomed so that they are approximately aligned (a given detection appears at the same place on both maps). All point-like 4XMM detections with log(likelihood) XX\_8\_DET\_ML $>$ 6.5 in the total band for the current instrument  (XX) are accumulated on each map. The MOS1 map is smoothed with a 3x3 boxcar average. The colour scale is square root between zero and three detections per pixel in MOS1, 0 and 100 in pn. Obvious hot areas are visible.  They appear in only one instrument because the detections on hot areas have DET\_ML > 6.5 only in the instrument where the hot area is, contrary to real detections. MOS2 is omitted because it shows no hot area in that part of the focal plane.}
\label{fig:map_detml}
\end{figure}

\begin{figure}
\centering
\includegraphics[width=\linewidth]{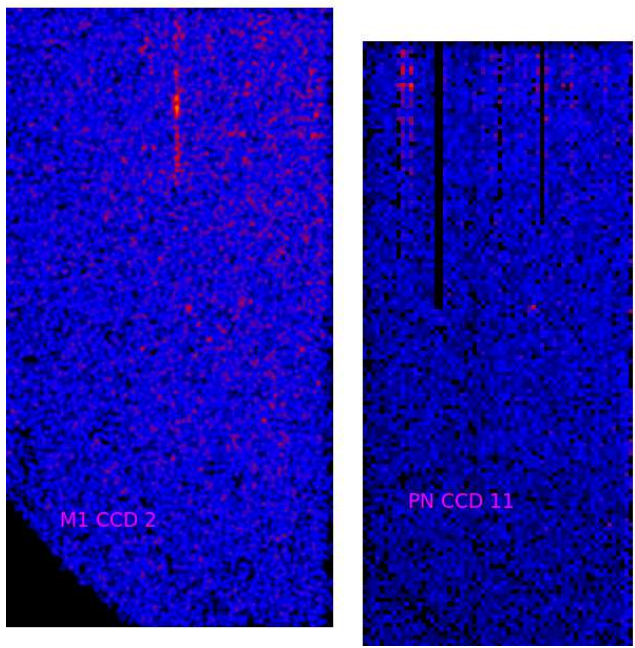}
\caption{Same region and colour scale as in Figure~\ref{fig:map_detml}. Detections on a hot area and inside the associated revolution interval are rejected. The numbers inside hot areas in which the revolution interval is not the full interval are corrected for the different time coverage. The remaining features cannot be distinguished from statistical fluctuations with the current algorithm.}
\label{fig:map_masked}
\end{figure}

Warm pixels on a CCD (at a few counts per exposure) are too faint to be detected as such by the automatic processing, but can either push faint detections above detection level, or create spurious detections when combined with statistical fluctuations. This is an intrinsically random process, not visible over a short period of time, but which creates hot areas when projecting all detections detected over 18 years onto the detector plane.

We addressed this by projecting for each CCD all detections onto chip coordinates (pn or M1 or M2\_RAWX or Y), keeping only detections above the detection threshold with the current instrument alone. In that way, we can distinguish hot areas coming from different instruments, see Figure~\ref{fig:map_detml}.

We proceeded to detect hot pixels or columns in each CCD, using a similar method to the SAS task embadpixfind. Because the localisation precision of faint detections is several arcseconds (larger than the MOS pixel size of 1.1$\arcsec$) the detection was carried out for MOS after binning the image to 3x3 pixels (and testing all 9 single-pixel shifts). We flagged hot pixels with a probability less than $10^{-2}/N_{\rm trials}$ to be compatible with a Poisson distribution at the local average (estimated from the local median plus 1). The trials factor $N_{\rm trials}$ was set to the image size (64x200) for pn and three times the binned image size (3x200$^2$) for MOS, accounting approximately for the fact that the shifted binned images are correlated.

Hot columns are detected in the same way after projecting the images (with hot pixels masked) onto RAWX. A column was considered bright when it was too high at the $7\sigma$ level applying the likelihood ratio test for Poisson counts \citep{LiMa83} with respect to its surroundings (excluding immediate neighbours). This very high threshold was chosen such that subtle increases not obvious by eye were not detected (there are hundreds of detections per column, so that method is very sensitive).

It often occurs that only a piece of a column is bright. In order to identify such occurrences we compared the distribution of detections along RAWY in the hot column with that in the same neighbouring columns used in the column detection, using the Kolmogorov-Smirnov (hereafter KS) test. If the probability of compatibility was less than $10^{-4}$, we looked for the bright interval with repeated KS tests on restricted lengths on each side of the RAWY value where the maximum distance between the two distributions occurs, until we reached a probability of compatibility larger than $10^{-2}$. The remainder on each side was considered normal or hot depending on the result of a Li \& Ma test at the $3\sigma$ level with respect to the neighbouring columns.

We defined contiguous hot areas after reprojecting all the hot pixels and segments of columns onto the CCD (at the full pixel resolution for MOS). Many of those warm pixels were not present at the beginning of the mission, and some appear for a short amount of time. So we tested each hot area for variability using revolution number, and the same KS-based algorithm used to detect segments of bright columns, compared to the reference established over all detections on all CCDs and all instruments. This resulted in a revolution interval for each hot area. The distribution of remaining detections is shown in Figure~\ref{fig:map_masked}.

Detections on a hot area for a particular instrument and within the corresponding revolution interval are flagged with flag 12. This results in 16,503 flagged sources for pn, 6,245 for MOS1 and 1,382 for MOS2.

\section{Source-specific product generation}
% - NW
\label{sec:SSP}

In order to minimise any contribution from soft proton flares, Good Time Interval (GTI) filtering is carried out. This is done for each exposure. A high 
energy light curve (from 7-15 keV for pn, $>$ 14 keV for MOS) is created,
and initial background flare GTIs are derived using the optimised approach employed in the SAS task,
bkgoptrate \citep{rose16}. Bkgoptrate determines the background count rate 
threshold at which the data below the threshold yields a maximum signal to noise ratio, by filtering the periods of time when the lightcurve count rate is above the optimised threshold. Following the identification of so called bad pixels, event cleaning and 
event merging from the different CCDs, an in-band (0.5-7.5 keV) image is then created, using
the initial GTIs to excise background flares. After source detection, an in-band light curve is generated, excluding events from 
circular regions of radius 60\arcsec\ for sources with count rates $\le$0.35
ct s$^{-1}$ or 100\arcsec\ for sources with count rates $>$0.35ct s$^{-1}$, centred on 
the detected sources. The SAS task, bkgoptrate, is then applied to the light curve to find the 
optimum background rate cut threshold and this is subsequently used to define the 
final background flare GTIs. If no lightcurve can be generated, a general filtering for the observation is carried out. Image data are extracted from events using GTIs determined from when the pointing direction is within 3\arcmin\ of the nominal pointing position for the  observation. 

Following the source detection process, detections identified with at least 100 EPIC counts have their spectra extracted.  If the number of counts not flagged as 'bad' (in the sense adopted by 
{\em XSPEC}) is still greater than 100 counts, a spectrum and a time series are extracted using an aperture around the source whose radius is automatically determined to maximise the signal-to-noise of the source data. This is done with a curve-of-growth analysis, performed by the SAS task,  eregionanalyse. The algorithm then searches for a circular background region on the same CCD where the source is located, excluding regions where sources have been detected, as described in \cite{rose16}.  The exception is in the case when the source falls on the central CCD of a MOS observation in SmallWindow mode (PrimePartialW2 or 3). In that case the background is estimated from an annulus (inner radius of 5.5\arcmin\ and outer radius of 11\arcmin) centred on the source. The background is therefore estimated from the peripheral CCDs and the central CCD is completely excluded. For EPIC-pn sources, the algorithm avoids the same RAWY column as the source in order to exclude out-of-time events from the background estimation. The background region always has a radius larger than three pixels, otherwise no background is calculated. Response files (.rmf and .arf) are then created using the SAS tasks rmfgen and arfgen.

The pile-up is estimated as described in Section~\ref{sec:pileup} and written to the header. 

\subsection{Lightcurve generation}
% - NW
\label{sec:lcs}

Lightcurves are corrected using the SAS task, epiclccorr, to take into account events lost through inefficiencies due to vignetting, bad pixels, chip gaps, PSF and quantum efficiency, dead time, GTIs and exposure. epiclccorr also takes into account the background counts, using the background lightcurve, extracted over the identical duration as the source lightcurve.  The time bin size for the pn lightcurves was previously set to a minimum of ten seconds and could be as poorly sampled as tens of thousands of seconds for the faintest sources. To exploit the high time resolution and high throughput of the pn, for 4XMM we now extract the pn lightcurve such that each bin is 20 times the frame time, usually 1.46 s. The binning of the MOS data remains as it was for 3XMM.

\subsection{Variability characterisation}
% - NW
\label{sec:varchar}

As in 3XMM, the $\chi^2$ test was used to determine if a source is variable during a single observation. Variability was defined as P($\chi^2$) $\le$ 10$^{-5}$. We also gave the fractional variability, F$_{var}$, to provide the scale of the variability \citep{rose16}. These values are still provided in the 4XMM catalogue. The $\chi^2$ statistic can be applied to binned data sets where the
observed number of counts in a bin deviates from expectation approximately
following a Gaussian distribution. \cite{cash1979} showed
that when the number of counts per bin falls below $\sim$10-20, the $\chi^2$ statistic becomes inaccurate. Therefore, as the pn lightcurves are now binned to 20 $\times$ frame time (Section~\ref{sec:lcs}), these data are rebinned to contain 20 counts per bin before applying the variability tests. Future versions of the catalogue are expected to exploit the high time resolution of the pn lightcurves using a Kolmogorov-Smirnov test which can be carried out on finely binned data.

%we determine the variability using a Kolmogorov-Smirnov test. Sources with a Kolmogorov-Smirnov probability $\le 10^{-5}$ of being constant are flagged as variable. 

As in previous catalogue versions, we still provide columns with the minimum EPIC source flux (and error) and the maximum EPIC source flux (and error).  This allows the user to find sources variable between observations. Alternatively, the fluxes from each observation, along with the observation date can be seen directly as a table when querying a source on the catalogue server\footnote{\url{xmm-catalog.irap.omp.eu}}. Whilst the majority of sources do not vary in flux, the maximum variability in the catalogue is a factor  $\times$10$^5$ in flux (for example, V2134 Oph a low mass X-ray binary).

Variability between observations is also provided in the stacked catalogue, see Section~\ref{sec:stackedcat}.

\section{Screening} 
% - NW
\label{sec:screening}

Visual inspection of each detection in every observation that was included in 4XMM was carried out, as has been done for previous versions of the catalogue \citep{rose16}. The aim of the screening is to visually validate the new methodology employed in the pipeline, ensure that the pipeline processing has run correctly, and to flag detections that are likely to be spurious and that have not been automatically identified as possibly spurious in the pipeline processing. Whilst the source detection process is very robust, some spurious detections can still occur in the wings of the PSF of a bright source, in reflection arcs caused by a bright source outside of the field of view, in very extended diffuse emission in the field of view, or because of anomalous noise in a region of the detector, for example. The regions affected are masked and any detections in such regions are subsequently assigned a manual flag (flag 11) in the flag columns ( pn\_FLAG, M1\_FLAG, M2\_FLAG, EP\_FLAG). The fraction of the field of view that is masked is characterised by the observation class (OBS\_CLASS) parameter. The definition of the OBS\_CLASS parameter is given in the Table~\ref{tab:obsclasses}, along with the percentage of the catalogue (4XMM-DR9 and 3XMM-DR8 for comparison) with that particular OBS\_CLASS value.
 
\begin{table}[ht]
\begin{minipage}[ht]{\columnwidth}
\normalsize
\caption{4XMM observation classification. }
\label{tab:obsclasses}
%\begin{center}
\small
\centering
\renewcommand{\footnoterule}{} 
\tabcolsep 0mm
\begin{tabular}{
c @{\extracolsep{1mm}} c @{\extracolsep{1mm}} c @{\extracolsep{1mm}} 
c }
\hline \hline
%       &      & \multicolumn{4}{c}{Filters}       \\
OBS CLASS & masked fraction & \multicolumn{1}{c}{3XMM-DR8} &
\multicolumn{1}{c}{4XMM-DR9}  \\
%& window\footnote{footnote...} \\
\hline
0 &  bad area = 0\%               & 18\%  &  30\%\\
1 &  0\% $<$ bad area $<$ 0.1\%   & 17\%  &  30\%\\
2 &  0.1\% $<$ bad area $<$ 1\%   & 16\%  &  17\%\\
3 &  1\% $<$ bad area $<$ 10\%    & 26\%  &  13\%\\
4 &  10\% $<$ bad area $<$ 100\%  & 14\%  &  9\%\\
5 &  bad area = 100\%             &  4\%  &  1\%\\

\hline
\end{tabular}
OBS\_CLASS, is the observation class given in the first column, the
  percentage of the field considered problematic is in the second column and the percentage of fields that fall within each class for 3XMM-DR8 and 4XMM-DR9 are given in the third and fourth columns respectively.
\end{minipage}
\normalsize
%\end{center}
\end{table}

There has been a marked improvement in the reduction in the number of spurious detections within each observation from the 3XMM to the 4XMM catalogue. This can be seen in Table~\ref{tab:obsclasses} which gives the area of each observation containing spurious detections. Of these observations, 77\% have less than 1\% of the field containing spurious detections, compared to only 51\% in 3XMM-DR8. Only 1\% of the fields in 4XMM-DR9 have no good sources in the field of view, compared to four times this value in 3XMM-DR8. This clearly shows the improvement in source detection, primarily due to the new background methods employed in the pipeline for 4XMM.

% \begin{figure}
%   \centering
%   \includegraphics[width=9cm]{OBSCLASS_DR8vsDR9.eps}
%      \caption{Percentage of observations with the different OBS CLASS values (see Table~\ref{tab:obsclasses} for precise definitions). Blue squares show data from 3XMM-DR8 and red diamonds show data for 4XMM-DR9. }
%         \label{fig:obsclasses}
%   \end{figure}

In 3XMM, flag 12 was not officially used. In 4XMM it indicates whether the source maybe spurious due to being on or close to warm or flickering pixels identified through stacking all of the detections in the 4XMM catalogue (see Sec.~\ref{sec:flagging}).

\section{Catalogue construction}
% - MC
\label{sec:cat_contruct}

\subsection{Unique sources}

The 4XMM detection catalogue contains multiple detections (up to 69 times in the most extreme case) of many X-ray sources, due to partial overlap between fields of view as well as repeated observations of the same targets. As has been done in previous versions of the catalogue \citep{rose16}, we assign a common unique source identifier, SRCID, to individual detections that are considered to be associated with the same X-ray source on the sky. The procedure used to perform associations is the same (and therefore subject to the same caveats) as the one outlined in section 6 of  \citet{rose16}. 

\subsection{Naming convention for the DETID and the SRCID}

Starting in 3XMM-DR5, the procedure for attributing the detection identification number (DETID)  and the unique source identification number (SRCID), both being unique to each detection and each unique source respectively, has been modified. Previously, identification numbers were re-computed for each catalogue version leading to supplementary columns added to the catalogue with the DETID and SRCID from previous releases. 

The DETID is now constructed from the OBS\_ID, which always remains the same for an observation, coupled with the source number SRC\_NUM\footnote{SRC\_NUM is the source number in the individual source list for a given observation; Sources are numbered in decreasing order of count rate (that is, the brightest source has SRC\_NUM = 1).} as follow:

\begin{equation*}
\mathrm{DETID} = ``1" + \mathrm{OBS\_ID} + \mathrm{SRC\_NUM} 
\end{equation*}
where the ``+'' sign indicates string concatenation and where SRC\_NUM is zero-padded to form a four digit number. The SRCID of a unique source is then determined from the first DETID attributed to that source (that is, in the observation where the source was first detected) and replacing the first digit ``1'' by ``2''.

Despite the new naming convention that aims at preserving SRCID numbers across catalogue versions, a certain number of SRCID can disappear from one catalogue version to another. This is a normal consequence of the algorithm that groups detections together into unique sources \citep[see section 6 of][]{rose16}. When new data are added and statistics are improved, the algorithm might find a better association of detections into unique sources. As an example, a total of 134 SRCIDs listed in 3XMM-DR7 are absent in 3XMM-DR8. 

\subsection{Missing detections and DETID change}

\begin{figure}
   \centering
   \includegraphics[width=9.2cm]{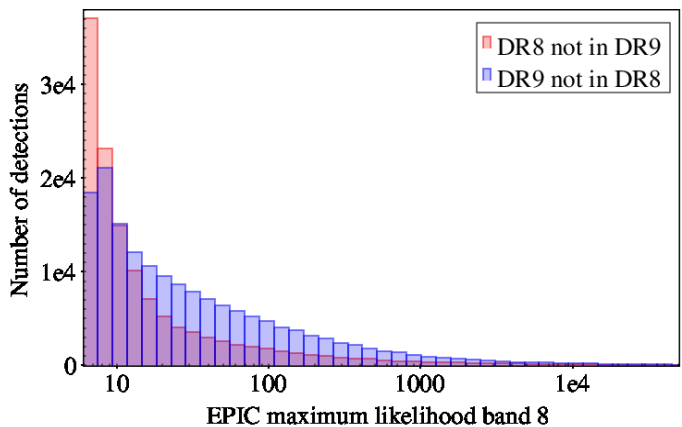}
      \caption{Histogram showing the detections present in 3XMM-DR8 and not present in 4XMM-DR9 as a function of maximum likelihood (red) and those in 4XMM-DR9 and not in 3XMM-DR8 (blue).}
         \label{fig:missing_sources}
   \end{figure}

In addition, the pipeline reprocessing of the full public dataset for the 4XMM version of the source catalogue led to significant modifications of the detection list. There are 10 214 observations that are common between the 3XMM-DR8 and 4XMM-DR9 catalogues, resulting in 773 241 detections in 3XMM-DR8 and 726 279 detections in 4XMM-DR9. Of these, there are 608 071 point-like detections with a SUM\_FLAG $\leqslant$ 1 in 3XMM-DR8 and 607 196 in 4XMM-DR9. However, amongst these observations, there are $\sim$ 128 000 detections that appear in 3XMM-DR8 that are not matched with a detection in the same observation in the 4XMM-DR9 catalogue within a 99.73\% confidence region (that is, 2.27 $\times$ POSERR). About 67 000 of these were classified as the cleanest (SUM\_FLAG $\leqslant$ 1), point-like sources in 3XMM-DR8 -- these are referred to as missing 4XMM detections in what follows. Conversely, there are $\sim$ 164 000 detections in the 4XMM-DR9 catalogue that are in common observations but not matched with a detection in 3XMM-DR8 within 99.73\% confidence region, approximately 107 000 of which are classed as being clean and point-like.
This is an expected consequence of the reprocessing which was already encountered in the transition from 2XMM to 3XMM \citep[see Section 8 and Appendix D in ][]{rose16}. The number of missing 4XMM detections is consistent with the number of missing 3XMM detections, where there were $\sim$25700 good detections that appeared in 2XMMi-DR3 that were not matched with a detection in the same observation in the 3XMM-DR5 catalogue \citep{rose16}. This amounts to $\sim$4.5\% which is of the same order as the number of missing sources in 4XMM (8.3\%). The origin of these source discrepancies between the two catalogues are the improvements made to the pipeline and in particular the new background estimation. The majority of the detections present in 3XMM-DR8 that are not present in 4XMM-DR9 are from the lowest maximum likelihoods, see Figure~\ref{fig:missing_sources}. A small change in the parameters can cause a source with a maximum likelihood close to the cut-off of 6, but none the less slightly above, to have a value slightly below the cut-off and therefore be excluded from the catalogue.  Conversely, the changes in the pipeline for sources just below the maximum likelihood cut-off of six and therefore not in 3XMM-DR8 can mean that they will then have a higher maximum likelihood and be present in 4XMM-DR9. As discussed in Section~\ref{sec:screening}, fewer obviously spurious detections are found in 4XMM-DR9 than in 3XMM-DR8, which is also reflected in Figure~\ref{fig:missing_sources}, where the detections found in 4XMM-DR9 and not in 3XMM-DR8 are generally more reliable (higher maximum likelihood).

A related consequence is that the source numbering within a given observation (that is, the SRC\_NUM) has been altered in 4XMM-DR9 by the detections added and those removed. Therefore, amongst the detections that are matched between 3XMM-DR8 and 4XMM-DR9, the majority of them have different DETIDs in 4XMM-DR9 and 3XMM-DR8 (since the DETID is constructed from SRC\_NUM). To minimise this effect, for the detections matched between the two catalogues, we have chosen to keep the original 3XMM-DR8 DETIDs instead of the newly generated ones for 4XMM-DR9. However, in doing so, we ended up with $\sim$ 36 000 DETID duplicates due to unmatched 4XMM-DR9 detections having the same DETID as matched 3XMM-DR8 detections. In such cases, we added 5000 to the DETID of the unmatched detection to create a new unique DETID. 

\subsection{New and revised data columns in 4XMM}
% - MC
We have taken the opportunity of this major release version to revise some data columns and introduce new ones to the catalogues of detections and unique sources (the slim version).  A pileup evaluation per instrument for each detection is now provided as three new columns: pn\_PILEUP, M1\_PILEUP, and M2\_PILEUP, see Section \ref{sec:pileup}. In 3XMM-DR8 and earlier versions, the extent likelihood EP\_EXTENT\_ML was provided only for sources detected as extended. We now provide the extent likelihood for all sources, see Section~\ref{sec:extentML}. The source extent of unique sources (SC\_EXTENT) is now calculated using a weighted average. We now provide the error on the total band extent of a unique source: SC\_EXT\_ERR. It is calculated in the same way as the errors on the other unique source parameters (for example,  the SC\_EP\_FLUX\_ERR or the SC\_HRn\_ERR) namely, as the error on the weighted mean:
\begin{equation*}
\mathrm{\rm SC\_EXT\_ERR} = \sqrt{\frac{1}{\sum_{i} \frac{1}{\rm EP\_EXTENT\_ERR_{i}^2}}} 
\end{equation*}
where EP\_EXTENT\_ERR$_{i}$ is the total band error on the extent of the $i^{th}$ detection of the unique source.

\subsubsection{Pile up information}
% - JB
\label{sec:pileup}

As of 4XMM we provide three new columns (PN\_PILEUP, M1\_PILEUP, and M2\_PILEUP) quantifying whether each detection may be affected by pile-up in any instrument. A value below unity corresponds to negligible pile-up (less than a few \% flux loss) while values larger than ten denote heavy pile-up. Pile-up is dependent on time for variable detections. We neglect that here, but we note that a variable detection is more piled-up than a constant one for the same average count rate, so our pile-up level can be viewed as a lower limit. We also neglect the slight dependence on the detection spectrum due to the event grade dependence of pile-up.

Our pile-up levels are not based on a fit of the full images using a pile-up model \citep{Pileup99}. For point sources, they are equal to the measured count rates reported in the catalogue over the full energy band, transformed into counts per frame, and divided by the pile-up threshold. The thresholds (at which the pile-up level is set to 1) are set to 1.3 counts per frame for MOS and 0.15 counts per frame for pn \citep{Jethwa2015}.

For extended sources, the pile-up level is equal to the measured counts per frame per CCD pixel at the source position  divided by the pile-up threshold, and therefore refers to the peak brightness, assuming this can be considered uniform at the pixel scale (4.1$\arcsec$ for pn). The threshold is set for all instruments to $5 \times 10^{-3}$ cts per frame per pixel, such that the flux loss is also a few \% when the pile-up level is 1.

Among 733,796 point detections, 1,171 have PN\_PILEUP $>$ 1, among which most (1,042) have SUM\_FLAG = 1, and only 30 are not flagged (SUM\_FLAG = 0). Only 68 detections have PN\_PILEUP $>$ 10, among which three are not flagged, all of them in Small Window mode. Similarly, 1,388 detections have M1\_PILEUP $>$ 1 (22 not flagged) and 1,458 have M2\_PILEUP $>$ 1 (25 not flagged). All the 167 detections with PILEUP $>$ 10 in any MOS are flagged. The large pile-up values are of course strongly correlated between instruments, and when both are in Full Window mode, MOS is slightly more piled-up than pn (the median ratio of MOS to PN\_PILEUP is 1.27). Overall the number of point-like detections with PILEUP $>$ 1 in any instrument is 2,042 (50 not flagged).

\subsubsection{Extent likelihood}
% - IT
\label{sec:extentML}
%The discrepancy in EP EXTENT ML for point like and extended sources

 All detections are tested for their potential spatial extent during the
  fitting process. The instrumental point-spread function (PSF) is convolved with a
  $\beta$ extent model, fitted to the detection, and the extent likelihood
  EP\_EXTENT\_ML is calculated as described by Section 4.4.4 of \citet{wats09}. A source is classified  as extended if its core radius (of the $\beta$-model of the PSF), $r_c > 6$\arcsec\ and if the extended model improved the likelihood with respect to the point source fit such that it exceeded a threshold of $L_{ext,min}$=4.  In the 4XMM
  catalogues, EP\_EXTENT\_ML is included for all detections, while it was
  set to undefined for point-like detections in previous catalogues. $L_{ext,min} \geq$4 indicates that a source is probably extended, whilst
  negative values indicate a clear preference of the point-like over the
  extended fit. As in the previous catalogue, a minimum likelihood difference
  of four has been chosen to mark a detection as extended. This threshold
  makes sure that the improvement of the extended over the point-like fit is
  not only due to statistical fluctuations but from a more precise description
  of the source profile.

\section{The stacked catalogue}
% - IT
\label{sec:stackedcat}

A second independent catalogue is compiled in parallel by the XMM-Newton SSC, called 4XMM-DR9s, where the letter 's' stands for stacked. This catalogue lists source detection results on overlapping XMM-Newton observations. The construction of the first version of such a catalogue, 3XMM-DR7s, is described in \citet{trau19}. The construction of 4XMM-DR9s essentially follows the ideas and strategies described there with important changes that are described in full detail in the accompanying paper Traulsen et al. (submitted). The two main changes concern the choice of input observations and event-based astrometric corrections before source detection. Also it was found necessary to perform some visual screening of the detections, whose results are reported in the source catalogue.

Observations entering 3XMM-DR7s were filtered rather strictly. Only observations with OBS\_CLASS$ < 2$, with all three cameras in full-frame mode, and with an overlap area of at least 20\% of the usable area were included. All those limitations were relaxed for the construction  of 4XMM-DR9s which resulted in a much larger number of observations to be included and potentially much larger stacks (more contributing observations). Before performing simultaneous source detection on the overlapping observations, individual events were shifted in position using the results from the previous catcorr positional rectification of the whole image processed for 4XMM-DR9. This led to a clear improvement of the positional accuracy in stacked source detection. 

All sources found by stacked source detection are listed in 4XMM-DR9s, including those from image areas where only one observation contributes. One may expect some differences between these same sources in 4XMM-DR9 and DR9s, because their input events were treated differently. More information is given in Traulsen et al. (submitted).

4XMM-DR9s is based on 1329 stacks (or groups) with 6604 contributing observations. Most of the stacks are composed of two observations, the largest has 352. The catalogue contains 288191 sources, of which 218283 have several contributing observations. Auxiliary data products comprise X-ray and optical images, and long term X-ray light curves. Thanks to the stacking process, fainter objects can be detected and 4XMM-DR9s contains more sources  compared to the same fields in 4XMM-DR9.

\section{Catalogue properties}
% - NW
\label{sec:properties}

\begin{figure}
   \centering
   \includegraphics[width=9.5cm]{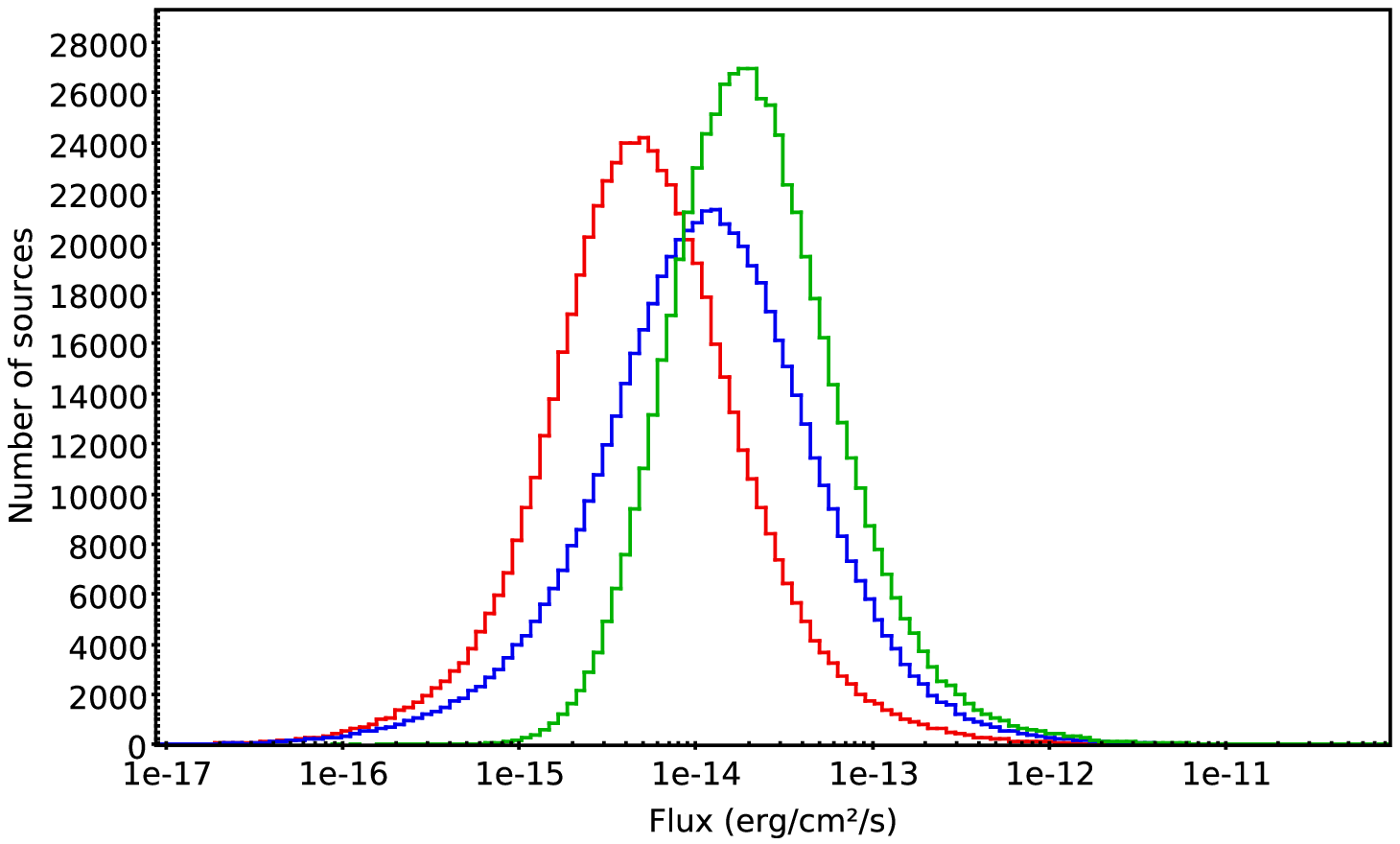}
   \hspace*{0.15cm}\includegraphics[width=9.3cm]{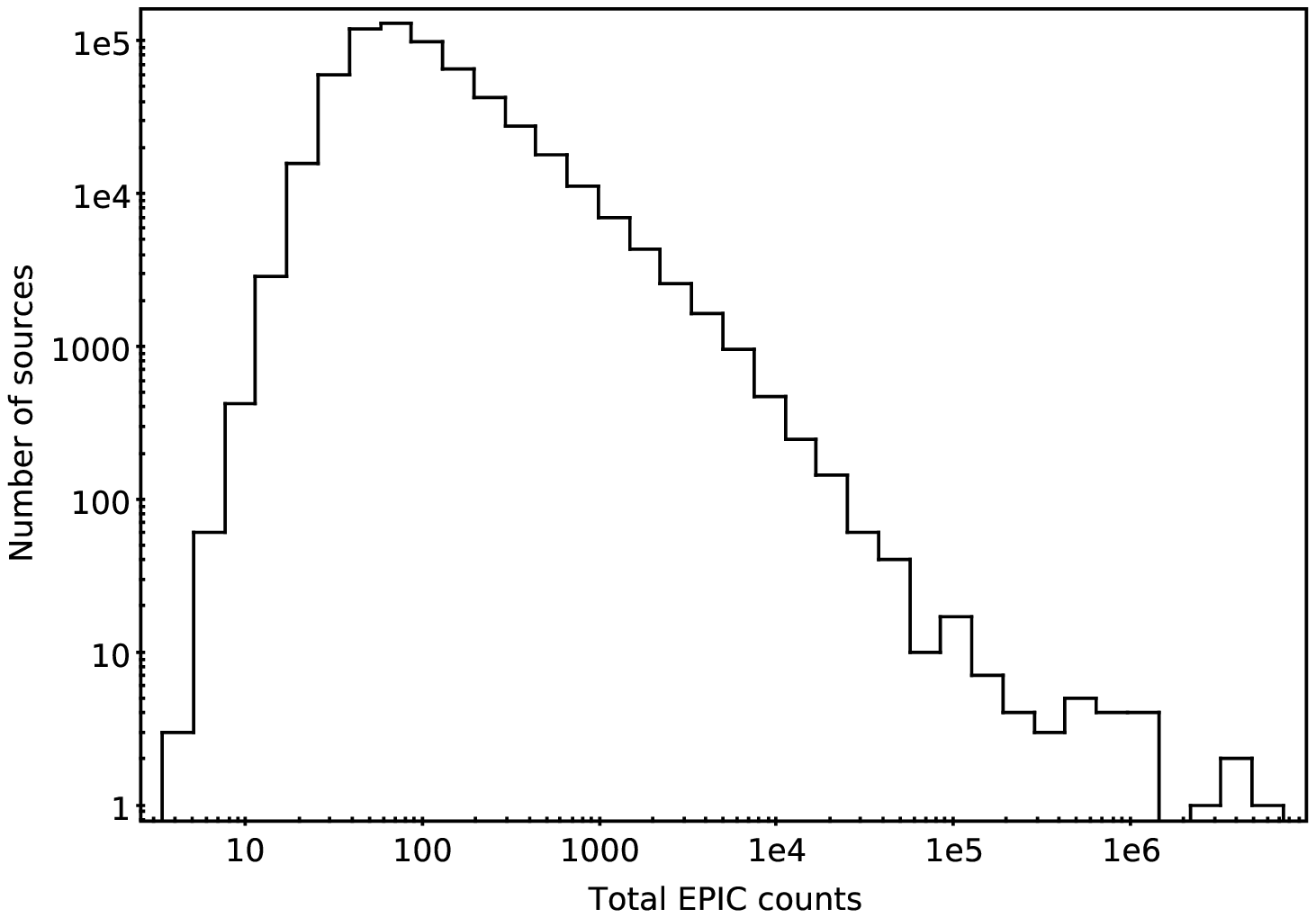}
      \caption{Top: Distribution of source fluxes for the 4XMM-DR9 catalogue in the soft (0.2-2.0 keV, red), hard (2.0-12.0 keV, blue), and total band (green) energy bands. Only sources with summary flag 0 are included. Bottom: distribution of total EPIC counts for the same sample of 4XMM-DR9 detections.}
         \label{fig:flux_dists}
   \end{figure}

The 4XMM-DR9 catalogue contains 810795 detections, associated with
550124 unique sources on the sky, extracted from 11204 public 
{\em XMM-Newton} observations. Figure~\ref{fig:flux_dists} shows the distribution of the source fluxes in the total EPIC band and in the soft and the hard band. Also shown in the figure is the distribution of the EPIC counts.

\begin{figure}
   \centering
   \includegraphics[width=9.5cm]{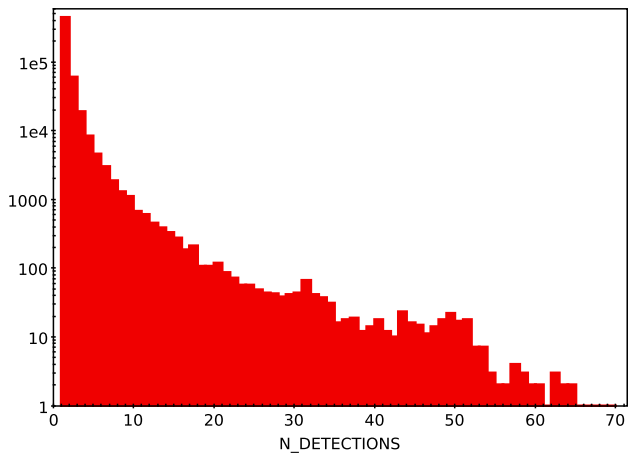}
      \caption{4XMM-DR9 unique sources plotted as a function of the number of detections}
         \label{fig:repeat_dets}
   \end{figure}

Amongst the 4XMM-DR9 detections, 121792 unique sources have multiple
detections, the maximum number of repeat
detections being 69, see Fig.~\ref{fig:repeat_dets}. 76999
X-ray detections in 4XMM-DR9 are identified as extended objects, that is, with a
core radius parameter,
$r_{c}$, as defined in section 4.4.4 of \cite{wats09}, $>$ 6\arcsec and EP\_EXTENT\_ML>=4, with 74163 of these having $r_{c} <$80\arcsec.

\subsection{Astrometry}
% - FC/MC 
\label{sec:astrometry}

 The systematic astrometric uncertainty of the 4XMM~DR9 detection catalogue has been estimated empirically using the SDSS~DR14~QSO catalogue \citep{Paris18}, following similar steps as those detailed in Section~\ref{sec:poserr}. However, here we use all of the detections in 4XMM-DR9 and any value of poscorrok. The sources in the SDSS~DR14~QSO catalogue have been filtered (good quality spectra and avoiding mismatches between targeting and matched photometry\footnote{see \url{https://www.sdss.org/dr15/spectro/caveats/}}). The filtered catalogue has then been cross-matched with the SDSS~DR9 photometry catalogue with a maximum distance of 5~arcsec. We have discarded all QSOs with more than one SDSS~DR9 counterpart out to that distance, keeping only pointlike objects (\texttt{cl=6}). We cross-correlated the 402291 ``clean'' quasars with the ``slim'' catalogue out to a distance of $r=30$\arcsec. For each of the resulting pairs we have estimated the combined positional error as in Section~\ref{sec:poserr} and discarded all quasars that had more than one counterpart out to $x=r/\sigma=6$, making 11640 suitable quasars (there were 43 quasars with more than one counterpart out to that limit).

\begin{figure}
\centerline{\includegraphics[width=0.49\textwidth]{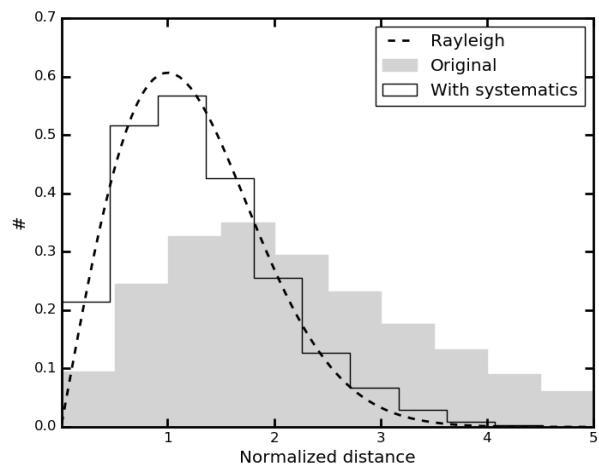}}
\caption{Fraction of XMM-Newton-SDSS quasar pairs as a function of normalised distance $x$, before adding a systematic uncertainty (grey histogram) and after its addition (black solid line), along with the Rayleigh distribution (black dashed line).}
\label{fig:rayleigh}
\end{figure}

Filtering as described in Section~\ref{sec:poserr} leaves 15001 quasar-X-ray pairs with $x<5$. To follow the Rayleigh distribution $P(x)=xe^{-x^2/2}$, we have added an additional positional uncertainty $\Sigma$ in quadrature, so that the total positional uncertainty is now $\sigma ' =\sqrt{\sigma^2+\Sigma^2}$, looking for the value of $\Sigma$ that minimises the log-likelihood of the $x' \equiv r/\sigma'$ and the Rayleigh distribution. We find $\Sigma=0.961\pm 0.008$~arcsec for the uncorrected 4XMM-DR9 X-ray positions, where the uncertainty (1$\sigma$) has been calculated by bootstrap with replacement. This can be seen in Figure~\ref{fig:rayleigh}.

\begin{figure}
\centerline{\includegraphics[width=0.49\textwidth]{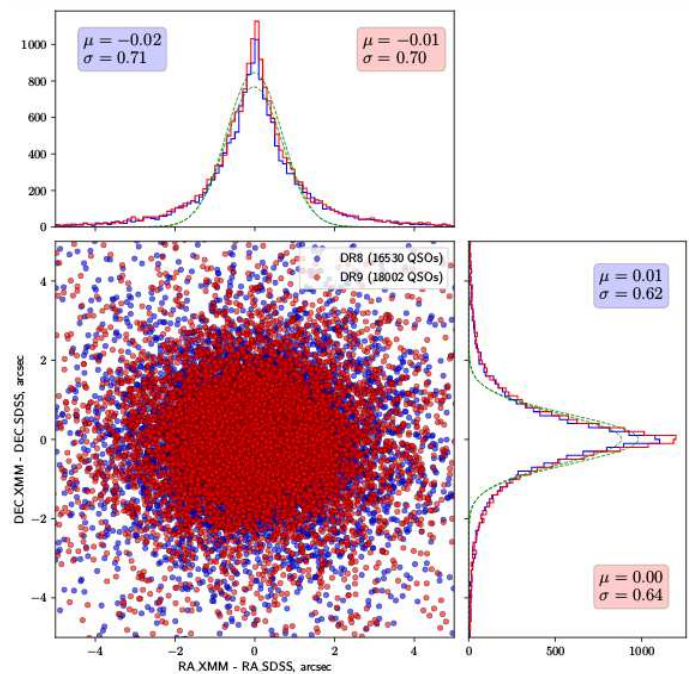}}
\caption{Scatter plot and associated distribution of the RA and Dec offsets between the XMM sources and the SDSS optical quasars. Two versions of the XMM catalogues are compared: 4XMM-DR9 (red) and 3XMM-DR8 (blue). The dashed green curves in the histogram plots represent gaussian fits to the distributions. The derived mean $\mu$ and standard deviation $\sigma$ for each fit are shown in the coloured boxes respectively.}
\label{astrometry}
\end{figure}

To directly compare the quality of the astrometry in 3XMM-DR8 and 4XMM-DR9, we matched each catalogue of detections with the DR14 release of the SDSS quasar catalogue. Cross-matching was performed without restrictions on the types of XMM-Newton and SDSS sources considered, but we kept only those matches within a matching radius of 15\arcsec. This yielded a total of 16530 3XMM-QSO pairs and 18002 4XMM-QSO pairs. Figure \ref{astrometry} shows a scatter plot and associated histograms of the RA and Dec offsets between the XMM sources and SDSS quasars. We see that the general astrometric quality of the 4XMM-DR9 catalogue is very good, with mean RA and Dec offsets of -0.01\arcsec\ and 0.005\arcsec\ respectively with corresponding standard deviation of 0.70\arcsec\ and 0.64\arcsec. No significant improvement is observed when comparing with the 3XMM-DR8 - SDSS match. 

\subsection{Extended sources}
% - NW
\label{sec: extended}

Only 76999 4XMM-DR9 detections (9.50\%) are identified as extended, compared to 91111 in 3XMM-DR8 (11.75\% of the catalogue). However, of the extended sources in 4XMM-DR9, 30464 have the best quality flag (SUM\_FLAG=0, 40\% of extended sources), whereas only 12256 of the 3XMM-DR8 extended sources (13\%) have this flag. This implies that the detection of extended sources is more reliable in the new version of the catalogue, with fewer spurious extended sources. This is due to the improved background modelling used for 4XMM-DR9.

\section{External catalogue cross-correlation}
% - CM
\label{sec:catxcorr}
Cross-correlation with archival catalogues is performed by a distinct pipeline module running at the Observatoire Astronomique de Strasbourg and referred to as the Astronomical Catalogue Data Subsystem (ACDS). For each individual EPIC detection the ACDS lists all possible multi-wavelength identifications located within a 3$\sigma$ combined XMM and catalogue error radius from the EPIC position. Finding charts and overlays with ROSAT all-sky survey images of the field are also produced. A detailed description of the ACDS is given in \cite{rose16}. 

We took the opportunity of the reprocessing of the entire XMM-Newton archive to update the list of archival catalogues and image servers entering the cross-correlation process and finding chart generation. In ACDS version 10.0, a total of 222 catalogues are queried, of which 53 are new with respect to ACDS version 9.0. Among the catalogues providing the largest sky coverage are;  GALEX GR6+7 \citep{bianchi2017}, UCAC4, SDSS DR12 \citep{sdss12}, panStarrs-DR1 \citep{panstarssdr1}, IPHAS DR2 \citep{iphasdr2}, Gaia DR2 \citep{gaiadr2}, 2MASS, AllWISE, Akari, NVSS, FIRST and GLEAM \citep{gleam}. The XMM-OM Serendipitous Source Survey Catalogue XMM-SUSS4.1 \citep{suss}, XMM-Newton slew survey Source Catalogue v. 2.0, the 3XMM-DR8 catalogues, Chandra V2.0 catalogue and the second ROSAT all-sky survey are also queried. Apart from the Chandra Catalogue Release 2.0 whose entries are extracted from the CXC server, all other ACDS catalogues are queried using the Vizier catalogue server.   

As for previous releases, 4XMM ACDS tentative identifications are not part of the catalogue proper but are distributed to the community by the XSA and through the XCAT-DB \citep{michel2015}\footnote{\url{http://xcatdb.unistra.fr}}.
Finding charts are extracted from several imaging surveys with the following decreasing priority order. First the Sloan digital sky survey \citep{sdss12} with colour images made from the $g$, $r$ and $i$ images extracted from the SDSS server. Second the Pan-STARRS-DR1 \citep{panstarssdr1} with colours images based on the $z$, $g$ and $z$+$g$ surveys, third, the MAMA/SRC-J and MAMA/POSS-E plate collections and as a last choice the DSS2 photographic plates. For the one colour photographic surveys, we select the blue image at Galactic latitude $>$ 20$^\circ$, while the red images are preferred in the Galactic plane. Apart from the SDSS, all images are extracted in HEALPix format from Hierarchical Progressive Surveys (HiPS) Aladin server \citep{hips2014}. 

\subsection{Methodology for producing multi-wavelength Spectral Energy Distributions}
% - CM/LM
Spectral energy distributions (SEDs) are provided for each of the unresolved (EP\_EXTENT=0) unique 4XMM sources. For that purpose, we use basically the same tools as those developed in the framework of the ARCHES project \citep{Arches2017}. The ARCHES algorithm \citep{pineau2017} cross-matches in a single pass all selected archival catalogues and for each combination of catalogue entries, computes the cross-match probability. Probabilities are computed from the likelihood that sources in the different catalogues have exactly the same position on the sky, considering their astrometric uncertainties. In particular, the resemblance of the derived SED with that of any given class of objects does not enter in the computation of the probability. The association probability eventually rests on the prior probability that a given X-ray source has a true counterpart in the longer wavelength catalogue considered. This prior is estimated from the observed distribution of X-ray - longer wavelength catalogue associations taking into account the expected rate of spurious matches. In the original ARCHES project, X-ray sources were grouped by XMM observations with similar exposure times, corresponding to similar limiting sensitivities. Although this grouping method offers a clean and relatively easy way to build X-ray source instalments, it still has the disadvantage of mixing bright and faint X-ray sources that will not have the same a priori probability to have a counterpart in the longer wavelength catalogues considered. In order to cope with this potential statistical bias, we designed a method aimed at grouping X-ray sources by range of X-ray flux instead. Accordingly, the ARCHES cross-matching tool had to be modified so as to read the sky area covered by the sample as an input instead of computing it from the list of observations given in entry. 

Source detection area requires building EPIC sensitivity maps for each of the XMM observations. In order to compute sensitivity maps, we first tried the approach proposed by for example, \cite{carrera2007}. The method consists of equating the probability of existence of a given source as provided by  EP\_8\_DET\_ML with that derived from an excess of counts above a given background assuming Poisson statistics. Although good fits can be obtained for EP\_8\_DET\_ML higher than $\sim$ 15, we found that best fit background areas are highly dependent on off-axis angle and background values when approaching the threshold of EP\_8\_DET\_ML = 6, used as a criteria for a detection to be included in the 4XMM catalogue. Such a discrepancy is not unexpected since the existence probabilities given by the emldetect algorithm also depend on the resemblance of the distribution of photons to that of the PSF. In addition, emldetect relies on the Cash statistics \citep{cash1979} and on the approximation of the Wilks theorem to derive probabilities. Instead, we built sensitivity maps by computing at each pixel location the total EPIC broad band count rate that would yield a mathematical expectation of EP\_8\_DET\_ML equal to 6. For that purpose we assume a power law input source spectrum ($\Gamma$ = 1.42; N$_{\mathrm H}$ = 1.7$\times10^{20}\,$cm$^{-2}$) similar to that of the unresolved sources contributing to the extragalactic background \citep{lumb2002}. The source spectrum is then folded through the exposure maps and filter responses so as to obtain the source counts in each band and camera in operation. EP\_8\_DET\_ML is then computed taking into account the background maps and the varying shape of the PSF with telescope and off-axis angle. 

We estimated the overlap of the 4XMM-DR9 catalogue with 26 archival catalogues selected to cover the largest sky coverage and widest span in wavelength from UV to radio. The Multi-Order-Coverage map (MOC) \citep{fernique2015} of each XMM observation was computed with a resolution of 12.8\arcsec\ (order 14) and compared to the MOC footprint of each catalogue using a python code developed at CDS \citep{baumann}. Table \ref{xmatchstat} lists the pre-selected catalogues sorted by 4XMM coverage. In the optical band, catalogues were prioritised according to their depth, astrometric quality, and range of colours in the following order, SDSS12, PanStarrs DR1 and Skymapper, so as to cover the entire sky. Whenever a GAIA DR2 match was found within 1.4\arcsec\ from the catalogue entry, the GAIA position was assigned to the merged source. APASS9 photometry was added to the merged source if found within a 1.4\arcsec\ distance so as to extend the photometric measurements to brighter objects. The 1.4\arcsec\ radius was derived from the shape of the Rayleigh distribution of the distances between matching sources and garantees a low rate of false cross-identification. In a similar manner, we cross-matched the ALLWISE and 2MASS catalogues keeping the 2MASS position whenever the difference of position was lower than 3.5\arcsec\ at $|b| \geq$ 20\,deg and 1.5\arcsec\ at $|b| \le$ 20\,deg. Special sky regions such as M31 and the LMC were discarded due to their high optical source density. For each unique source, we only kept the observation offering the highest detection area. 4XMM sources were then grouped into four EPIC (0.2-12.0 keV) ranges of flux with boundaries at 1.4,  
3.1 and 7.2$\times$ 10$^{-14}$ erg cm$^{-2}\,s^{-1}$. This grouping yields a nearly even number of sources in each flux band.

The statistical ARCHES cross-match procedure was applied to five catalogues or group of catalogues: XMM, Galex, SUSS-OM, merged optical and merged infrared. Due to the different areas of the non all-sky catalogues (Galex, SDSS12, PanStarrs and Skymapper) we split the XMM observations into groups having homogeneous catalogue coverages. In addition, the galactic plane region was treated separately. Finally, a simple cross-match between the ARCHES result and both the AKARI and merged FIRST/NVSS compiled by \cite{mingo2016} was made. However, their matching likelihoods do not enter in the computation of the overall SED probability provided by the ARCHES tool.

A standard table at CDS\footnote{\url{http://vizier.u-strasbg.fr/viz-bin/VizieR-3?-source=METAfltr}} allows us to convert magnitudes into flux. The resulting SEDs are available as individual FITS files and graphical output for the three highest probability SEDs. 

\begin{table}
\begin{center}
\caption{Overlapping area between photometric catalogues and 4XMM observations. The last column shows the way the catalogue was processed, either using the ARCHES multi-catalogue statistical cross-match (s) or using a simple positional cross-match (x)}
\label{xmatchstat}
\begin{tabular}{lcccc}
\hline
Catalogue            &  Total area    &  Overlap  & Band & Xmatch\\
                     &  covered       &  with    &      & mode  \\
		     &                & 4XMM     &      &        \\
		     &  (deg$^{2}$)   & (deg$^{2}$)  & &\\
\hline
AllWISE              &      all-sky   &  1152 & ir &s\\
Gaia DR2             &      all-sky   &  1152 & opt &s\\
UCAC4                &      all-sky   &  1152 & opt &\\
2MASS                &      all-sky   &  1152 & ir &s\\
APASS                &      all-sky   &  1126 & opt &s\\
Akari                &       39406   &  1108 & farir& x\\
GMRT                 &       36996   &  1000 & radio &\\
NVSS                 &       34069   &   927 & radio & x\\
PanStarrs DR1        &      32134   &   881  &opt & s\\
GalexGR67            &      26249   &   696  & uv & s\\
GLEAM                &       25423   &   657 & radio &\\
SkyMapper            &      19585   &   550  & opt & s\\
SDSS12               &      14520   &   504  & opt & s\\
FIRST                &       10847   &   427 & radio& x\\
VHS                  &      13670   &   364  & ir&\\
XMM-OM-SUSS41        &         348   &   343 &uv & s\\
SUMSS                &        8354   &   216 & radio &\\
UKIDSS LAS           &       3695   &   174 & ir &\\
VST                  &       3988   &    86 & opt &\\
Galex MIS            &       1880   &    83 & uv &\\
VPHAS                &         670   &    77 & opt &\\
UKIDSS GPS           &       1366   &    76  & ir & \\
WBH2005 20           &         614   &    72 & radio &\\
Glimpse              &         471   &    70 & ir & \\
IPHAS                &       1888   &    59  & opt &\\
WBH2005 6            &         164   &    35 & radio &\\
\hline
\end{tabular}
\end{center}              
\end{table}

The sensitivity maps, individual observation MOCs and total 4XMM MOCs are available on the XMM-SSC website\footnote{\url{http://xmmssc.irap.omp.eu/Catalogue/4XMM-DR9/4XMM_DR9.html}}.

%\section{Examples - NW}
%\label{sec:examples}

\section{Catalogue access}
% - NW
\label{sec:access}

% Give examples of how to create a clean catalogue
The catalogue of detections is provided in several formats. A Flexible Image
Transport System (FITS) file and a comma-separated values (CSV) file are
provided containing all of the detections in the catalogue. For 4XMM-DR9 there
are 810795 rows and 336 columns. A separate version of the catalogue (the slim
catalogue) with only the unique sources is provided, that is, 550124
rows, and has 45 columns, essentially those containing information
about the unique sources. This catalogue is also provided in FITS and CSV
format. We also provide SQL CREATE statements to load the data in CSV format. These can be found on the {\em XMM-Newton} Survey Science Centre
webpages\footnote{\url{http://xmmssc.irap.omp.eu/}}. The stacked catalogue is provided in FITS format only. Ancillary tables to the catalogue also available from the
{\em XMM-Newton} Survey Science Centre
webpages include the table of
observations incorporated in the catalogue.

The {\em XMM-Newton} Survey Science Centre webpages provide access to the 4XMM
catalogue, as well as links to the different servers distributing the full
range of catalogue products. These include, the ESA {\em XMM-Newton} archive (XSA), which
provides access to all of the 4XMM data products, and the ODF data, the
XCat-DB\footnote{\url{http://xcatdb.unistra.fr/4xmm/}} produced and maintained by
the {\em XMM-Newton} SSC, which contains possible EPIC source identification produced 
by the pipeline by querying 222 archival catalogues, see Section~\ref{sec:catxcorr}. Finding charts are also provided 
for these possible identifications. Other source properties as well as images, time series and
spectra are also provided. Multi-wavelength data 
taken as a part of the XID (X-ray identification project) run by the SSC over  the first
fifteen years of the mission are also provided in the XIDresult 
database\footnote{\url{http://xcatdb.unistra.fr/xidresult/}}. The XMM-SSC catalogue server\footnote{\url{http://xmm-catalog.irap.omp.eu/}} provides access to each source and regroups information concerning all of the detections for a unique source. It also provides the XMM-Newton lightcurves and spectra and permits the user to undertake simple spectral fitting, as well as overlays of the same region of sky in all wavelengths.  The catalogue can also be accessed through
HEASARC\footnote{\url{http://heasarc.gsfc.nasa.gov/db-perl/W3Browse/w3table.pl?tablehead=name\%3Dxmmssc\&Action=More+Options}}
and VIZIER\footnote{\url{http://vizier.u-strasbg.fr/cgi-bin/VizieR}}.  The results
of the external catalogue cross-correlation carried out for the 4XMM catalogue
(section~\ref{sec:catxcorr}) are available as data products within the XSA or through the XCat-DB.  The {\em XMM-Newton} Survey Science Centre webpages also detail how to provide feedback on the catalogue.

Where the 4XMM catalogue is used for research and publications, please 
acknowledge their use by citing this paper and including the following: \\

{\em This research has made use of data obtained from the 4XMM {\em XMM-Newton} serendipitous source catalogue compiled by the ten institutes of the {\em XMM-Newton} Survey Science Centre selected by ESA.}

We note that the 4XMM catalogue of detections, as for previous versions of this catalogue, contains detections with a significance as low as $\sim$3 $\sigma$ (Maximum likelihood of 6), along with sources that have been flagged as possibly spurious. Statistically some of these sources will be spurious. In order to create the cleanest catalogue possible, where statistically almost all sources are real, it is necessary to filter the catalogue to include only EPIC sources with for example, a 5 $\sigma$ significance (Maximum likelihood of $\sim$14) and to keep only those with with no flags, for example, 

{\tt EP\_8\_DET\_ML $>$ 14 \&\& SUM\_FLAG $<$ 1}

Filtering with these criteria for 4XMM-DR9 leaves 433612 detections. 99.6\% or 431924 of the point-like detections have no pileup (XX\_PILEUP $<$ 1, where XX is either pn, M1 or M2 for the pn, MOS 1 or the MOS 2 detectors).

\section{Upper limits for observed regions of the sky}
% - NW
\label{sec:upper_limits}

The XMM-SSC provides an upper limit server for the user to determine an upper limit for the flux given a non-detection in a region observed by {\it XMM-Newton}. The server is known as FLIX (Flux Limits from Images from XMM-Newton). This upper limit can be calculated for any of the standard {\it XMM-Newton} bands for a user defined statistical significance and sky region. A single region or many regions may be queried at the same time. This upper-limit flux is determined empirically using the algorithm described by \cite{carrera2007}. A link to the FLIX upper limit server is provided on the XMM-SSC webpages and the ESA SOC webpages\footnote{\url{https://www.cosmos.esa.int/web/xmm-newton/xsa}}.

\section{Limitations of the catalogue} 
% - NW
\label{sec:limitations}

\subsection{Maximum extent of extended detections}
%Why 80"
%  - NW/IT
When dealing with extended detections, the software determines the radius of the detection, up to a limit of 80\arcsec\ to optimise processing time. Whilst this may appear restrictive, only 0.007\% of the catalogue detections are clean and extended, with a radius of $>$80\arcsec.

\subsection{Error values on counts, rate and flux}
%  - NW
Should a detection fall close to a chip gap or the edge of the field of view on one or more cameras, only a small fraction of the point spread function will be recorded for that camera. The fraction is given by the XX\_MASKFRAC columns, where XX refers to EP (EPIC), PN (pn), M1 (MOS 1) or M2 (MOS 2). Where the XX\_MASKFRAC value is low, the error on the counts, rate or flux may be very high, compared to the value of the counts, rate or flux, as these quantities are derived for the whole PSF. We note that detections which have less than 0.15 of their PSF covered by the detector are considered as being not detected.

\section{Future catalogue updates}
\label{sec:updates}
%NW
Incremental releases (data releases) are planned to augment the 4XMM catalogue.  At least one additional year of data will be included with each data release. Data release ten (DR10) will provide data becoming public during 2019 and should be released during 2020. These catalogues will be accessible as described in Section~\ref{sec:access}.

\section{Summary}
% - NW
This paper describes the improvements made to the software and calibration used to produce the new major version of the {\it XMM-Newton} catalogue, 4XMM. 4XMM-DR9 contains 810795 detections in the X-ray band between 0.2 and 12.0 keV. The catalogue covers 1152 degrees$^2$ (2.85\%) of the sky. In terms of unique X-ray sources, the 4XMM-DR9 catalogue is the largest X-ray catalogue produced from a single X-ray observatory, with 550124 unique sources compared to 317167 unique X-ray sources in the Chandra source catalogue v. 2.0 and 206335 unique X-ray sources in the 2SXPS catalogue of X-ray sources from the Neil Gehrels Swift Observatory. In this new version of the catalogue, source detection has been shown to be much improved, with fewer spurious sources and in particular, many fewer spurious extended sources. In addition, we provide lightcurves and spectra for a much larger fraction of the catalogue than in previous versions (36\% of detections in 4XMM-DR9 compared to 22\% of detections in 3XMM-DR8). These spectra and lightcurves benefit from finer binning (MOS spectra and pn lightcurves). The catalogue benefits from extra complementary products, such as multi-wavelength spectral energy distributions for each source, sensitivity maps and catalogue footprint maps.  We provide information on how to access the catalogue as well as how to retrieve upper limits for non-detections in the catalogue footprint. The catalogue is ideal for quick access to data products (fluxes, spectra, images, etc), searching for new objects, population studies of homogenous samples and cross correlation for multi-wavelength studies.

\begin{acknowledgements}
We are grateful to the anonymous referee for careful reading of the manuscript and for providing useful feedback. We are grateful for the strong support provided by the XMM-Newton SOC. We also thank the CDS team for their active contribution and support. The French teams are grateful to Centre National d'\'Etudes Spatiales (CNES) for their outstanding support for the SSC activities.   SSC work at AIP has been supported by Deutsches Zentrum f\"ur Luft- und Raumfahrt (DLR) through grants 50OX1701 and 50OX1901, which is gratefully acknowledged.  FJC acknowledges financial support through grant AYA2015-64346-C2-1P (MINECO/FEDER). MTC and FJC acknowledge financial support from the Spanish Ministry MCIU under project RTI2018-096686-B-C21 (MCIU/AEI/FEDER/UE) cofunded by FEDER funds and from the Agencia Estatal de Investigaci\'on, Unidad de Excelencia Mar\'ia de Maeztu, ref. MDM-2017-0765.  This paper used data from the SDSS surveys.  This research has made use of the VizieR catalogue access tool, CDS, Strasbourg, France (DOI : 10.26093/cds/vizier). The original description  of the VizieR service was published in 2000, A\&AS 143, 23. This paper made use of the topcat software \citep{tayl05}.

\end{acknowledgements}

   \bibliographystyle{aa} % style aa.bst
   \bibliography{4XMMv7} % your references Yourfile.bib

%\begin{thebibliography}{}

%  \bibitem[Baker(1966)]{baker} Baker, N. 1966,
%      in Stellar Evolution,
%      ed.\ R. F. Stein,\& A. G. W. Cameron
%      (Plenum, New York) 333

%\end{thebibliography}
%\appendix %First appendix
%\section{Data modes of XMM-Newton exposures included in the 3XMM catalogue.}
%\label{ap:datamodes}

\end{document}